\documentclass[apj]{emulateapj}
\usepackage{epsfig}

\newcommand{\kpc}{\,\mbox{kpc}}
\newcommand{\kms}{\,\mbox{km}\,\mbox{s}^{-1}}

\newcommand{\msun}{\,M_{\sun}}

\newcommand{\tsim}{\sim\!}
\newcommand{\ea}{et al.}
\newcommand{\metals}{[\mbox{Fe}/\mbox{H}]}
\slugcomment{Submitted to Astrophysical Journal Letters}
\shorttitle{Was the Andromeda Stream Produced by a Disk Galaxy?}
\shortauthors{Fardal et al.}

\defcitealias{fardal06}{F06}
\defcitealias{fardal07}{F07}
\defcitealias{gilbert07}{G07}
\defcitealias{ibata07}{I07}

\begin{document}
\title
  {Was the Andromeda Stream Produced by a Disk Galaxy?}
\author{Mark A. Fardal\altaffilmark{1,2,3},
Arif Babul\altaffilmark{3},
Puragra Guhathakurta\altaffilmark{2},
Karoline M. Gilbert\altaffilmark{2},
Cara Dodge\altaffilmark{1,4}}
\altaffiltext{1}{
  Dept.\ of Astronomy, University of Massachusetts, 
  Amherst, MA 01003, USA}
\altaffiltext{2}{
  UCO/Lick Observatory, Dept.\ of Astronomy \& Astrophysics,
  Univ.\ of California, 1156 High St., Santa Cruz, CA 95064, USA}
\altaffiltext{3}{
  Dept.\ of Physics \& Astronomy, University of Victoria, 
  Elliott Building, 3800 Finnerty Rd., Victoria, BC, V8P 1A1, Canada}
\altaffiltext{4}{Astronomy Department, Smith College, Clark Science Center,
        Northampton, MA 01060, USA}
\email{fardal@astro.umass.edu}

\begin{abstract}
The halo region of M31 exhibits a startling level of stellar
inhomogeneities, the most prominent of which is the ``giant southern
stream''.  Our previous analysis indicates that this stream, as well
as several other observed features, are products of the tidal
disruption of a {\it single\/} satellite galaxy with stellar mass $\tsim
10^9 \msun$ less than 1~Gyr ago.  Here we show that the specific
observed morphology of the stream and halo debris favors
a cold, rotating, disk-like progenitor over a dynamically hot,
non-rotating one.  These observed characteristics include the
asymmetric distribution of stars along the stream cross-section and
its metal-rich core/metal-poor sheath structure.  We
find that a disk-like progenitor can also give rise to arc-like
features on the minor axis at certain orbital phases that resemble
the recently discovered minor-axis ``streams'', even
reproducing the lower observed metallicity of these streams.  Though
interpreted by the discoverers as new, independent tidal streams,
our analysis suggests 
that these minor-axis streams may alternatively arise from the progenitor
of the giant southern stream.  
Overall, our study points the way to a more complete reconstruction of
the stream progenitor and its merger with M31, based on the emerging
picture that most of the major inhomogeneities observed in the M31
halo share a common origin with the giant stream.
\end{abstract}

\section{INTRODUCTION}
The relative proximity of the Andromeda galaxy (M31) and the global perspective
from our external vantage point 
make M31 an excellent laboratory for studying the
stellar halos of large galaxies.  Resolved
stellar maps of M31's halo, assembled over the past decade, have revealed
highly complex inhomogeneities, the most striking
of which is the Giant Southern Stream (GSS), extending
$\tsim 150 \kpc$ away from
M31's center in the southeast direction \citep[][hereafter I07]{ibata01,
ferguson02,mcconnachie03,ibata07} and falling towards M31's center
with relative radial velocities as high as $\tsim 250 \kms$
\citep{ibata04,raja06,kalirai06}.
Other significant morphological and kinematic features
in the M31 halo include stellar shelves \citep[][hereafter
\citetalias{fardal07}/\citetalias{gilbert07}]{ferguson02,fardal07,gilbert07}
as well as the recently discovered minor axis ``streams''
\citepalias{ibata07}.  The GSS is
especially notable because it offers an opportunity to
precisely measure M31's potential
\citep[][hereafter \citetalias{fardal06}]{ibata04,fardal06} and provides a view
into the most significant Local Group galaxy disruption in the last Gyr.

\begin{figure*}[t!]
\begin{center}
\includegraphics[bb=5 0 535 348, width=6.5in]{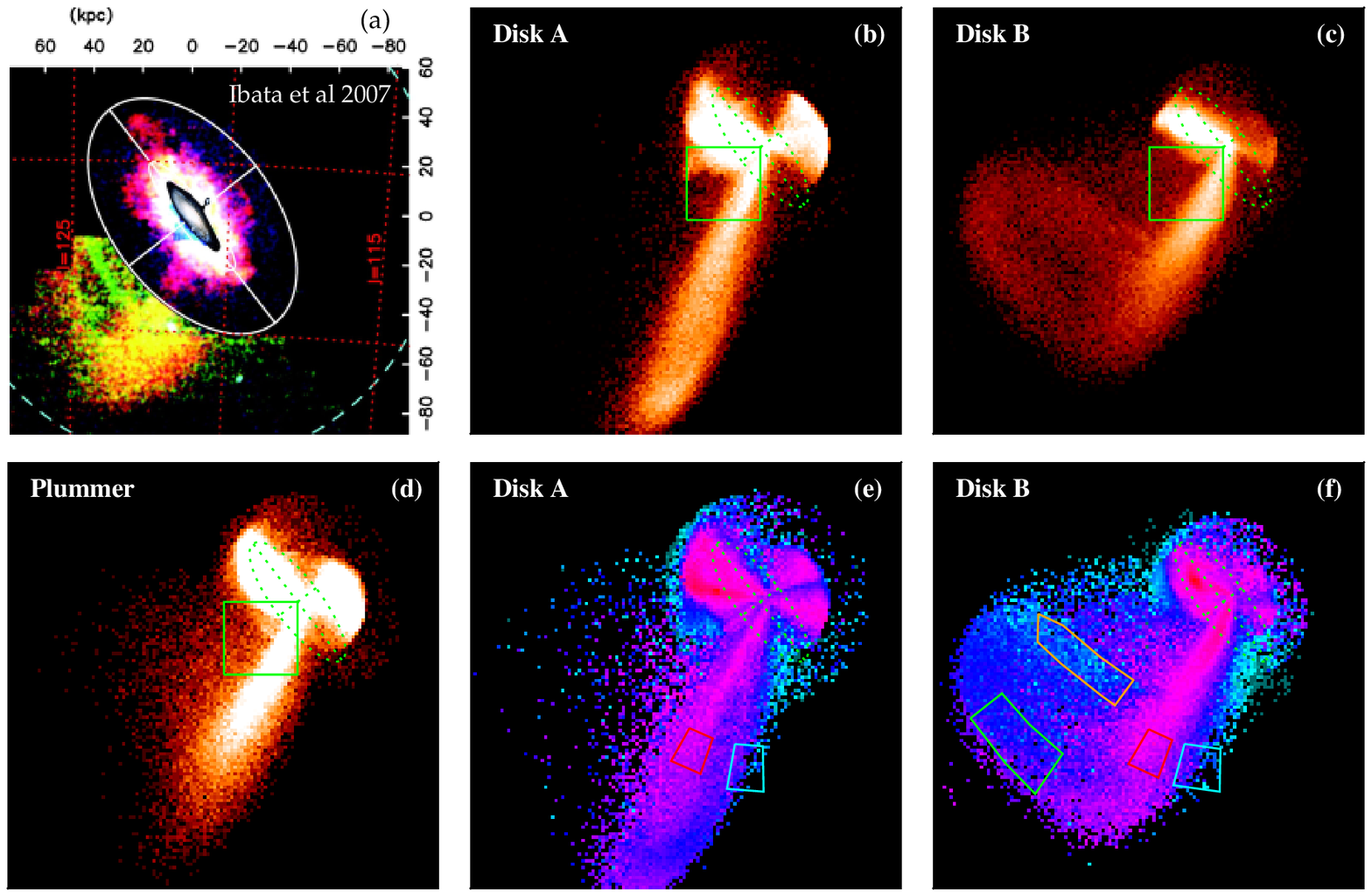}   
\caption{
\label{fig.skymaps}
(a): Stellar surface density/metallicity map of M31 from
\citetalias{ibata07}.
The shallower INT/WFC survey is used inside the large ellipse,
and the deeper CFHT/MegaCam survey outside it.
The minor-axis ``streams'' C and D are visible at the lower left,
projecting from the larger plume of the GSS.
These streams (green) and the GSS cocoon (red) are observed to be more
metal poor than the GSS core (yellow).
(b): Mass surface density map from model Disk~A,
840~Myr into the run.  The map is 160~kpc on a side, and a
dotted contour indicates M31's disk orientation.
The square indicates the region shown in Figure~\ref{fig.fieldbars}.
(c): Same for model Disk~B, at 680~Myr.
(d): Same for the Plummer model, at 840~Myr.
(e): Map of the metallicity as a function of position
in Disk~A (at 840~Myr), with red denoting the highest metallicity,
dark blue intermediate, and light blue the lowest (see 
Figure~\ref{fig.metals}a for a quantitative scale).
Boxes indicate the regions used for the metallicity
histograms in Figure~\ref{fig.metals}. 
(f): Same for Disk~B (at 680~Myr).
}
\end{center}
\end{figure*}

Models detailing the formation of the GSS agree remarkably well 
with most aspects of the observations, and suggest
the progenitor had a stellar mass of
$\tsim 2 \times 10^9 \msun$ \citep[][\citetalias{fardal06};
\citetalias{fardal07}]{font06}.  
Our kinematic analysis in F07 finds that seemingly unrelated features 
like the ``Northeast Shelf'' and less prominent ``Western Shelf'' are also the
result of the same disruption process \citepalias{fardal07}, a conclusion
supported by independent studies of their stellar populations 
\citep{ferguson05,richardson08}.  The observed
GSS's most striking point of contrast with the models is its
asymmetry in the transverse direction.  As shown both with photometric
samples \citep{mcconnachie03} and spectroscopic surveys
\citepalias{gilbert07}, its stellar distribution is sharply truncated on
the NE side and falls off much more slowly on the SW side.  In
addition, the current models do not address the
observed stellar population gradients within the GSS
\citep[][I07]{ferguson02,mcconthesis06}.

In this letter, we show that this structure in the GSS can be
accounted for if the progenitor hosted a cold, rotating stellar disk,
unlike the simple spherical progenitors used in previous simulations.
Surprisingly, we find that the disruption of a disk galaxy can also
give rise to features similar to the recently discovered arc-like
minor-axis ``streams'', leading to the tantalizing
possibility that most of the major inhomogeneities observed in the M31
halo are tidal debris from the same galaxy that caused the GSS.
In Section 2, we briefly describe our model for the progenitor and
our $N$-body study of its tidal disruption.  In Section 3, we
show results from these simulations, focusing on the transverse
density profile of the GSS, the metallicity gradient, and arc-like
structures that overlap the minor axis.  Section 4 summarizes our
conclusions.

\section{SIMULATION METHOD}
Our simulations are based on the methods worked out in our earlier papers:
\citet{geehan06}, \citetalias{fardal06}, and \citetalias{fardal07}.  We use
the orbit and potential from Table~1 of \citetalias{fardal07} and their
spherical Plummer model to represent a non-rotating
progenitor.  For runs with a disk progenitor, we use the same initial
position and velocity, but substitute a different initial structure
of the satellite.

Briefly, our disk models assume the satellite is composed of a bulge
and rotating disk of stars.  For simplicity we assume that the dark
matter associated with the galaxy has been tidally stripped before the
encounter modeled here.  We use a hot exponential $\mbox{sech}^2$ disk
with mass $1.8 \times 10^9 \msun$, radial scale length $0.8 \kpc$, and
vertical scale height $0.4 \kpc$.  We add to this a Hernquist bulge of
mass $4 \times 10^8 \msun$ and scale length $0.4 \kpc$.  We initialize
both components with the package ZENO written by Josh Barnes.  We
evolve the satellite in M31's potential starting from 12 evenly spaced
orientations of the disk.  From this we select two models displaying
particularly good agreement with observational features, referred to
here as Disks~A and B.  In a forthcoming paper we will conduct a more
systematic survey of possible initial conditions and quantify the
debris structure in detail (Fardal \ea, in preparation).

The remaining details of the simulations are the same as those in
\citetalias{fardal06} and \citetalias{fardal07}.  We set the satellite
in motion inbound and slightly past apocenter, minimizing initial
transients from M31's tidal forces.  We run the simulations with the
multistepping tree code PKDGRAV \citep{stadel01}.  Our simulations
include the self-gravity of the progenitor satellite, but ignore
dynamical friction, perturbations from the other M31 satellites,
and the history of the progenitor prior to the orbit that produces the
GSS, as justified in our earlier papers.  We stop
the simulations 840~Myr into the run, which is
approximately 650~Myr past the initial pericenter.  This time was
chosen in \citetalias{fardal07} to give a reasonable match to the
debris structure around M31, including the radii of the ``shells'' on
the E and W sides.

\section{RESULTS}
\subsection{Stream morphology}
Figures~\ref{fig.skymaps}b,d show surface density maps based
on the Disk~A and Plummer models, respectively.
Both models reproduce the main feature of a stream extending
to the SE.  They also reproduce the observed line-of-sight distances and
velocities along the GSS.
However, the transverse distribution of GSS stars is strikingly different
between the two models---Disk~A displays a much sharper NE edge.
The observed star-count maps \citep[][\citetalias{ibata07}]{ferguson02} 
are not directly comparable since they contain both non-GSS-related 
M31 components and non-M31 contaminants and are not explicitly calibrated 
to stellar surface density, but the morphology of the GSS in these 
maps appears much closer to our disk model.

\begin{figure}[ht!]
\includegraphics[width=3.0in]{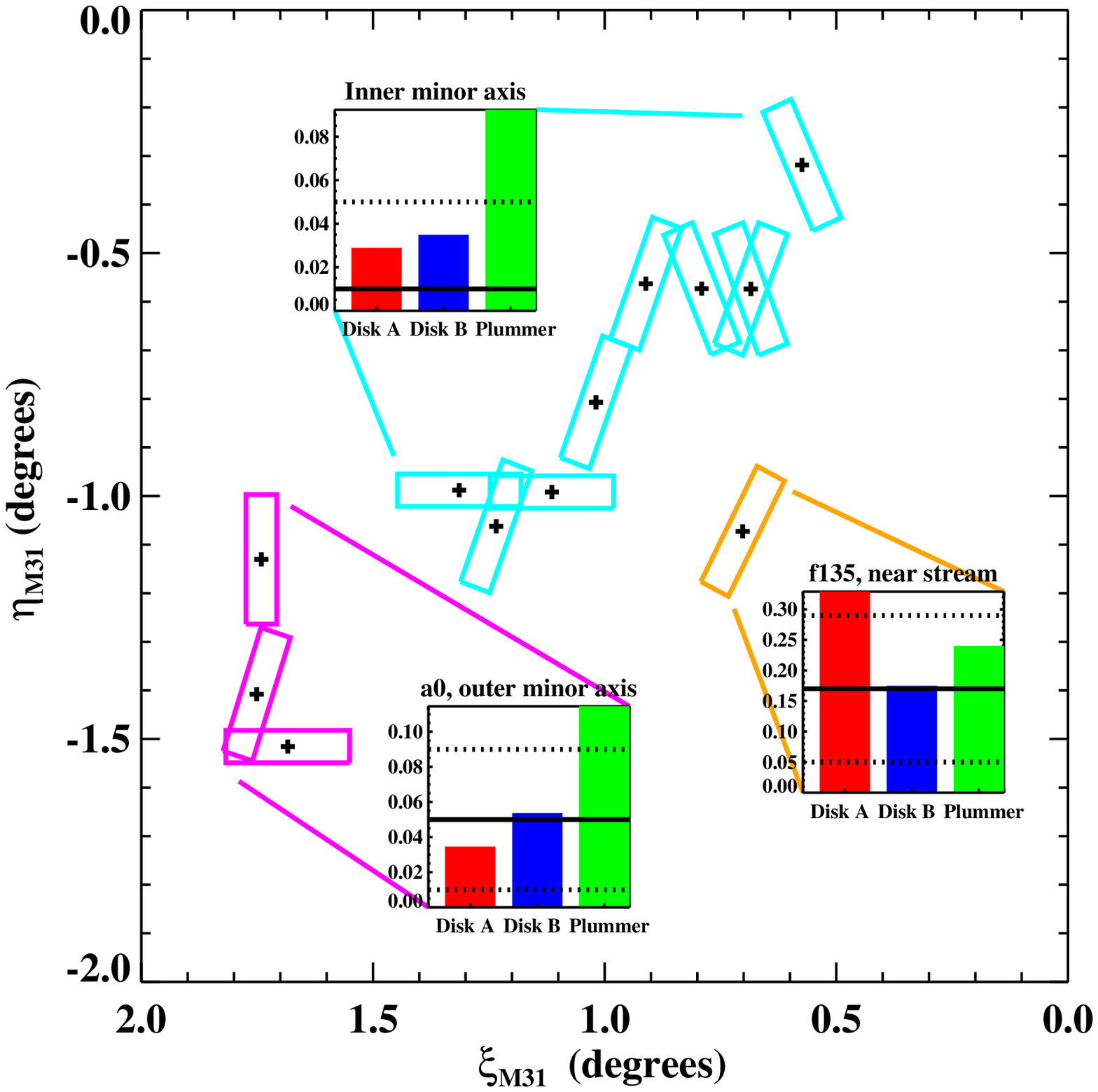}   
\caption{
\label{fig.fieldbars}
Comparison of the minor-axis contamination to the observations of 
\citetalias{gilbert07}.
The \citetalias{gilbert07} DEIMOS masks (rectangles) are grouped into
inner minor-axis masks, outer minor-axis masks, and a single
mask (f135) offset from the minor axis.  
M31's center is at (0,\,0).
The inset plots for each group show the 
ratio $R_m$ of the strength of the GSS component 
to the peak of the GSS at the same $R_{\rm proj}$.  
$R_m$ is measured as discussed in the text
for Disks A and B and the Plummer model 840~Myr into the runs.  The observational 
estimates and $\pm2\sigma$ error bars from Gilbert \ea\ (in preparation) are 
plotted as horizontal solid and dotted lines.  The Plummer model 
clearly contributes too much debris on the minor axis.
}
\end{figure}

Figure~\ref{fig.skymaps}d shows that the Plummer model results in a
large amount of stars spilling over as far as the SE minor axis,
located to the NE of the GSS.  When \citetalias{gilbert07} compared
their Keck/DEIMOS spectroscopic data near M31's SE minor axis to this model,
they noted much less spillover from the GSS than predicted by
the model.  Gilbert \ea\ (in preparation) has quantified this by dividing the
number of stars moving with GSS-like velocities on the minor axis to
those in the GSS core at the same projected radius $R_{\rm proj}$.  For the nine
innermost DEIMOS masks on the minor axis combined, 
this ratio $R_m = 0.01 \pm 0.02$; for
the three outermost masks on the minor axis, $R_m = 0.05 \pm 0.02$; and
for the mask f135 located somewhat nearer the GSS, they find a
likely detection of GSS material with $R_m = 0.17 \pm 0.06$.  

In Figure~\ref{fig.fieldbars}, we compare the density of GSS
stars in all three $N$-body models to these results.  We have selected
``GSS'' particles by defining the trend of radial velocity $v$ with
$R_{\rm proj}$ and then taking stars that fall within
$\pm 80 \kms$ of this velocity in the given field.  We also restrict
the particles to those actually in the GSS's ``shell''. 
We then repeat the procedure for a control field
located at the peak of the GSS at the same $R_{\rm proj}$, using
a smaller interval $\pm 40 \kms$ as the GSS core has a sharper
velocity distribution.  Clearly
the two disk models are in better accord with the observations than the
Plummer model.

The sharper NE edge and smaller minor-axis contamination of the disk
models thus imply that the progenitor was rapidly rotating.  We will
explore this argument in more detail in Fardal \ea\ (2008, in preparation).

\subsection{Metallicity pattern}
\begin{figure*}[ht!]
\begin{center}
\includegraphics[bb=10 0 537 177, width=6.5in]{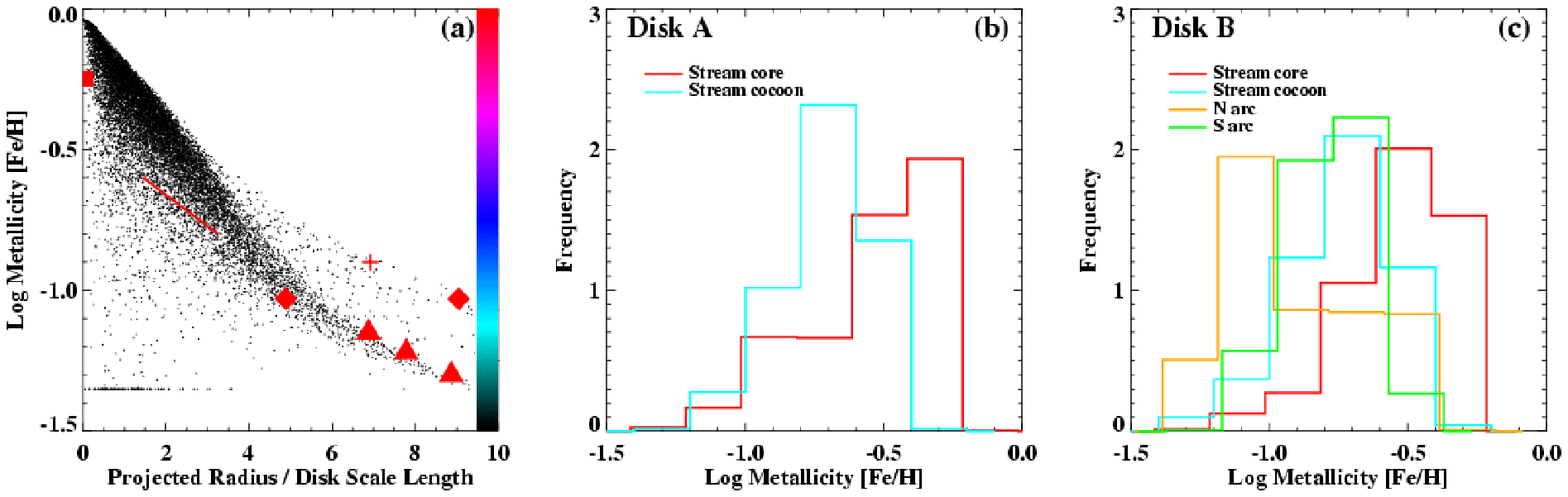}   
\caption{
\label{fig.metals}
(a): Metallicity values in the original disk prior to disruption
are shown by black dots, where the radius is plotted in units 
of disk scale length.
For comparison, observed results for M33's disk stars are plotted as 
symbols and lines: 
\citet{stephens02} (square);
linear approximation to points of \citet{kim02} (straight line);
\citet{galleti04} (diamonds);
\citet{mcconnachie06} (cross);
\citet{barker07} (triangles).
The colorbar translates $\metals$ to the color scale of 
Figures~\ref{fig.skymaps}ef.
(b): Histogram of particle metallicity values in model Disk~A within the 
``core'' and ``cocoon'' regions marked in Figure~\ref{fig.skymaps}e.  
(c): Histogram of particle metallicity values in model Disk~B within the 
``core'', ``cocoon'', ``N arc'', and ``S arc'' regions
marked in Figure~\ref{fig.skymaps}f.  
}
\end{center}
\end{figure*}

The mean color of GSS RGB stars is observed to vary
in the transverse direction: the GSS is significantly broader
in blue than in red stars \citep{ferguson02,mcconthesis06}.  This
is probably due to a metallicity gradient.  \citetalias{ibata07} quantified
the metallicity distribution in two GSS-dominated regions, one in
the center of the GSS and one in a less dense ``cocoon'' region to
the SW, and showed that the latter has a lower mean
metallicity.

Disk galaxies, of course, tend to
have metallicity gradients.  Therefore it is interesting to see how a
plausible gradient in our disk progenitor translates to the
metallicity pattern on the sky.

We use a simple parametric model to produce a plausible metallicity
gradient in our initial disk model.  We first find the specific
orbital energy $E_i$ of each particle.  We then assign it
a metallicity using
$\metals = A_Z + B_Z \, \log_{10} \left[ -E_i/(50 \kms)^2 \right]$,
setting $A_Z$ and $B_Z$ to agree with the results of \citetalias{ibata07}
as explained below.  This produces the metallicity gradient seen in
Figure~\ref{fig.metals}a.  Observational results for the stars in the
small disk galaxy M33 are also plotted, with the radius for both
galaxies normalized by the disk scale length; the metallicity pattern
of our disk model agrees quite well.  The GSS progenitor should
perhaps be lower in metallicity than M33 by a few tenths of dex due to
its lower inferred mass, but the photometric metallicity measurements
probably have systematic uncertainties at this level in any case.

Figure~\ref{fig.skymaps}e shows the sky view of the resulting
model metallicity pattern.  The gradient along the stream
is very weak, but the mean metallicity along the denser central part
is clearly higher than in the broad wing to the SW, similar to the
pattern seen in M31's GSS.  Using
\citetalias{ibata07}'s Figure~27,
we estimate the ``core'' and ``cocoon'' regions (at $R_{\rm proj}\sim60$~kpc)
have mean
metallicities of $\metals = -0.54$ and $-0.71$, respectively (mean
$\metals=-0.51$ was obtained at the GSS' sharp NE edge
by \citealp{raja06}).  Figure \ref{fig.skymaps}e shows ``broad wing''
and ``central GSS'' boxes chosen at a similar radius, but better
matching the slightly different model stream position.  Once we set $A_Z =
-0.70$ and $B_Z = 1.06$, the metallicities in these boxes are also
$-0.54$ and $-0.71$.  
The bare fact we can match two metallicities with two parameters is
not in itself meaningful, but it is significant that the magnitude
{\em and sign} of our initial metal gradient are very reasonable
(Fig.~\ref{fig.metals}a).
Figure~\ref{fig.metals}b shows that within each box there is a wide
range of metallicities; the distributions in \citetalias{ibata07}
appear somewhat broader, but given measurement errors and the
contributions from other halo components this is not surprising.

\subsection{Minor-axis arcs}
Using their MegaCam photometric survey of M31's halo, \citetalias{ibata07}
found multiple surface density ridges 
along the minor axis which they called ``streams''.
Streams C and D (the two closest to M31) form a pair
of curving ridges at slightly different orientations, which appear to
merge as they approach the survey boundary (see their Fig.~22).
Stream C appears to be slightly broader than stream D, and slightly
more metal-rich, though not as metal-rich as the GSS core/cocoon.
From I07's Figure~33 we estimate the mean metallicity of
streams C and D to be $-0.82$ and $-0.91$ respectively.
\citet{mori08} suggested these ``streams'' might be shell features 
from a satellite disruption, similar to the event
that created the GSS but from a different progenitor.

While studying our overall sample of runs based on 12 disk orientations, we
noticed one (Disk~B) containing two curious ``arcs'' crossing the minor axis.
These arcs are clearest at the step 680~Myr into the
run shown in Figure~\ref{fig.skymaps}c.  Morphologically, the two arcs somewhat
resemble streams C and D, with a fatter southern arc nearly merging
into a sharper northern arc.  Like the observed ``streams'', neither
arc crosses the GSS to the SW.  Compared to the observed arcs, the
simulated arcs are significantly further from M31's center.  

As Figure~\ref{fig.skymaps}f shows, the simulated arcs are significantly
less rich in metals than the GSS.  Using the same metallicity
model as for Disk~A and the regions defined by boxes in this figure,
the mean $\metals$ is $-0.78$ for the southern arc and $-0.90$ for the
northern arc.  Thus there is considerable if inconclusive evidence
that these arcs are close analogues of the ``streams'' in
\citetalias{ibata07}.

In our model, these two arcs originate from the outer regions of
the disk, and are sharp mainly because of the relatively cold velocity
field of the disk.  Both arcs consist of material that takes a path
around M31's center nearly opposite to the bulk of the progenitor,
explaining why they lie so far from the GSS.  The large size
of our disk is thus crucial; a compact progenitor
resembling M32, for example, would be unable to produce similar arcs.
The southern arc consists of a group of particles sharing nearly the same
energy, and come from fairly far out in the progenitor's disk.  The
northern arc consists of particles that lie even further out (explaining
its lower metallicity on average), which form a tidal tail during the
interaction with M31.

We cannot yet explore the full parameter space of the collision for
the presence and properties of these arc-like features. However, we
did conduct a few additional runs with changes to the disk mass,
radius, and orientation of Disk B, finding the arcs were sensitive to
the exact input parameters.  Thus we will require more theoretical
investigation as well as more observational constraints to determine
whether the arcs explain some of the \citetalias{ibata07} minor-axis
streams, or are merely a fortuitous similarity.
If the arcs are shown to be related
to the GSS, they will be a very solid argument for the disk nature of
the progenitor, and will place strong constraints on the parameters of
the collision.

\section{CONCLUSIONS}
In summary, several strands of observational evidence suggest that the
GSS originated from a progenitor with a strong sense of rotation, such
as a disk galaxy.  The transverse density
profile of the GSS is more easily produced by a rotating
satellite.  The observed decline in mean metallicity from the central core
of the GSS to its ``cocoon'' to the SW suggests that the progenitor had a
strong radial metallicity gradient, of the sort found mainly in disk
galaxies.
Furthermore, several observed arcs lying across the minor axis in M31
have very suggestive analogues in one of our runs.  If shown to be
related to the GSS in the manner suggested by our model, these
features would be definite confirmation of a disk-like progenitor.

The notion of a disk galaxy progenitor is somewhat at odds with age
measurements of the GSS, which suggests little star formation
during the last 4 Gyr \citep{brown06a,brown06b}.  However,
the fields used to infer this were placed in the central, metal-rich part of the
GSS; it is possible that the progenitor had an age gradient as well
as a metallicity gradient, with the older stars on the inside.
Age measurements in the GSS cocoon would therefore be
interesting.  It is also possible that the GSS progenitor was more similar 
to an S0 galaxy than a spiral, perhaps due to stripping of its
gas in an earlier phase of its encounter with M31.

Many papers have used metallicity to assess
the relationship among various M31 disk and halo features.  Our
suggestion that the GSS progenitor had a strong metallicity gradient
means that metallicity can no longer be used as a
reliable fingerprint of origin.  This obviously complicates the forensic
reconstruction of M31's merger history.  Despite this, the rapidly
growing database on M31 halo structure is a fascinating
puzzle, offering unique insight into the life of a typical disk
galaxy and the death of its unfortunate former companions.

\vskip+2mm

We thank Tom Quinn and Joachim Stadel for the use of PKDGRAV, Josh
Barnes for the use of ZENO, and Alan McConnachie and Roger Davies for
helpful conversations.
{}

\end{document}